\documentclass[draft]{agujournal2019}
\usepackage{url} 
\usepackage{soul}
\draftfalse

\journalname{Geophysical Research Letters}

\begin{document}

%
%


\title{Asymmetry of AMOC Hysteresis in a State-of-the-Art Global Climate Model}

%
%




\authors{Ren\'e M. van Westen\affil{1}, Henk A. Dijkstra\affil{1,2}}


\affiliation{1}{Institute for Marine and Atmospheric research Utrecht, Department of Physics,  Utrecht University, Princetonplein 5,3584 CC Utrecht, the Netherlands}
\affiliation{2}{Centre for Complex Systems Studies, Department of Physics, Utrecht University, Leuvenlaan 4, 3584 CE Utrecht,  the Netherlands}




\correspondingauthor{Ren\'e M. van Westen}{r.m.vanwesten@uu.nl}



\begin{keypoints}
\item A wide AMOC hysteresis is found  in a state-of-the-art global climate model 
\item The AMOC recovery is about six times faster than the AMOC collapse
\item The North Atlantic sea-ice cover plays a dominant role in the hysteresis asymmetry 
\end{keypoints}

%
%

%
%


\begin{abstract}
We study hysteresis properties of the  Atlantic Meridional Overturning Circulation 
(AMOC) under a slowly-varying North Atlantic (20$^{\circ}$N -- 50$^{\circ}$N)  freshwater 
flux forcing in state-of-the-art Global Climate Model (GCM), the Community Earth 
System Model. Results are presented  of a full  hysteresis simulation ($4,400$ model 
years) and show that there is a hysteresis  width of about $0.4$ Sv.  This demonstrates 
that an AMOC collapse and recovery  do not only occur in conceptual  and idealised 
climate models, but also  in a state-of-the-art GCM.  The AMOC recovery is about a factor
six faster than the AMOC collapse and this asymmetry is due to the major effect  of the 
North Atlantic sea-ice distribution on the AMOC recovery. The results have implications for projections 
of possible future AMOC behaviour and for explaining relatively rapid climate transitions 
in the geological past.  
\end{abstract}

\section*{Plain Language Summary}
The  Atlantic Meridional Overturning Circulation (AMOC) is considered to be a tipping element 
in the climate system.  We here simulate AMOC tipping events by slowly varying the North Atlantic 
freshwater forcing  in a state-of-the-art  Global Climate Model. The AMOC collapses when this 
forcing is sufficiently large and, when reversing this forcing,   the recovery occurs at much smaller 
values of the forcing than the collapse, giving rise to hysteresis behaviour.   The AMOC changes 
are much faster during its recovery than during its collapse  and this asymmetry is caused by the 
effect of the North Atlantic sea-ice distribution on the AMOC recovery.  The results demonstrate  
that AMOC tipping does not only occur in simplified climate models, but in  a large  hierarchy of 
climate models. 

%
%

%


%
%
%
%

\section{Introduction}

The future development of the Atlantic Meridional Overturning Circulation (AMOC) under 
climate change is a hot topic of research \cite{Ditlevsen2023}. The AMOC, as part of the 
global ocean circulation is crucial in maintaining the meridional heat transport in Earth's climate 
\cite{Srokosz2015}.   Models of the Climate Model Intercomparison Project  phase 
6 (CMIP6) project that the AMOC  strength will gradually decrease by the end of this 
century, with relatively  little differences between the various Shared Socioeconomic 
Pathway  scenarios \cite{Weijer2020}.  However, the AMOC has also been recognised 
as a climate tipping element \cite{Armstrong2022}, with a possible abrupt change in 
AMOC strength having severe global impacts \cite{Orihuela2022}.  Strong fluctuations 
in AMOC strength have occurred in the geological past, for example during the so-called 
Dansgaard-Oeschger  events, leading to local temperature variations of more than 
10$^{\circ}$C on Greenland \cite{Lynch-Stieglitz2017}.  

First ideas on AMOC tipping were proposed already in the 1960s \cite{Stommel1961}
using a highly conceptual model of the AMOC in which the ocean circulation was only 
represented by the flow between two boxes.  In this conceptual view, the present-day 
AMOC state is sensitive to changes to  its North Atlantic surface freshwater flux due to 
the so-called salt-advection feedback. Freshwater anomalies in the North Atlantic will decrease
the strength of the AMOC, thereby decreasing the northward salt transport and hence amplifying 
the original freshwater anomaly \cite{Marotzke2000}. In conceptual models, AMOC tipping is 
induced  by increasing the freshwater flux forcing and transitions occur between multiple  
equilibrium  AMOC states which exist for the  same freshwater forcing conditions 
\cite{Cessi1994}.

In more detailed climate models, there is evidence of multiple AMOC equilibrium states 
due to the appearance of  hysteresis behaviour.  When the freshwater forcing in the North 
Atlantic is increased very slowly, such that the AMOC remains in near-equilibrium, the 
AMOC is found \cite{Rahmstorf2005}  to collapse  in  many Earth system Models of 
Intermediate Complexity (EMICs).  When reversing the freshwater forcing,  the AMOC 
recovery occurs at much smaller freshwater forcing strengths than when it collapsed, 
giving rise to hysteresis behaviour. Such hysteresis behaviour has been found in relatively 
coarse climate models, such as the FAMOUS model \cite{Hawkins2011} and in an early 
version of the Community Climate System Model  \cite{Hu2012}.  

However, in state-of-the-art Global Climate Models (GCMs) the computation of such hysteresis behaviour
is very costly and so far only transient experiments with large freshwater perturbations 
have been performed to force AMOC collapses \cite{Weijer2019, Jackson2022}.  Substantially 
weaker (or collapsed) AMOC states have been found in these transient simulations 
\cite{Stouffer2006, Mecking2016, Jackson2018a}, but it has been difficult to identify a 
freshwater forcing regime where hysteresis behaviour occurs  \cite{Jackson2018b}. Hence, 
one might conclude  that AMOC tipping does not occur in state-of-the-art GCMs, because 
these models  include many more (in particular,  negative)  feedbacks than idealized 
climate models such as  EMICs and FAMOUS \cite[]{Gent2018}. 

Here we show results from a full hysteresis numerical simulation with a state-of-the-art GCM, 
the Community Earth System Model (CESM), using a configuration that is shortly described in 
Section~2.  The results 
in Section 3 show a large hysteresis width with  a strong asymmetry in the collapse and recovery 
of the AMOC. The analysis of these results, in particular the asymmetry of the hysteresis, show 
the important role for sea ice in the AMOC recovery. Implications of the results are discussed in 
Section~4. 

\section{Methods}

We use the CESM version~1.0.5 (the f19\_g16 configuration)  with horizontal resolutions of  1$^{\circ}$ 
for the  ocean/sea ice  and 2$^{\circ}$ for the atmosphere/land components.  The  simulation was 
branched off from the end (model year~2,800) of the  pre-industrial CESM control simulation described in 
\citeA{Baatsen2020}. There it is  shown that  the upper 1,000~m of the ocean is well equilibrated after 
2,800~years of model integration. We first linearly increased  the surface freshwater forcing  between 
latitudes   20$^{\circ}$N and 50$^{\circ}$N (inset in Figure~\ref{fig:Figure_1}a) with a rate $3 \times 
10^{-4}$~Sv~  yr$^{-1}$ up to model year 2,200, reaching a freshwater flux forcing of $F_H = 0.66$~Sv. 
This freshwater flux anomaly is compensated over the rest of the domain to conserve salinity. 
Starting from  the end of that simulation (model year~2,200)  we linearly decreased the freshwater 
flux forcing back to zero with a rate of  $-3 \times 10^{-4}$~Sv~ yr$^{-1}$. The rate of change is
comparable to that in \citeA{Hu2012}, who used $2 \times 10^{-4}$~Sv~ yr$^{-1}$, also  over 
the same  area in the North Atlantic.  

In the results below, we use the following diagnostics. The AMOC strength is defined as the 
total meridional volume transport at 26$^{\circ}$N over the upper 1,000~m:
 \begin{equation} \label{eq:AMOC}
\mathrm{AMOC}(y = 26^{\circ}\mathrm{N}) = \int_{-1000}^{0} \int_{x_W}^{x_E} v~\mathrm{d}x \mathrm{d}z
 \end{equation}
The freshwater transport by the overturning (AMOC)  component ($F_{\mathrm{ov}}$) and azonal (gyre) 
component ($F_{\mathrm{az}}$)  as a function of the meridional coordinate $y$ are determined as: 
\begin{eqnarray}
F_\mathrm{ov}(y)   &= & - \frac{1}{S_0} \int_{-H}^0 \left[ \int_{x_W}^{x_E} v^* \mathrm{d} x \right] \left[ \langle S \rangle - S_0 \right] \mathrm{d}z \\
F_\mathrm{az}(y)  &= & - \frac{1}{S_0} \int_{-H}^0 \int_{x_W}^{x_E} v' S' \mathrm{d} x \mathrm{d}z 
\end{eqnarray}
where $S_0 = 35$~g~kg$^{-1}$ is a reference salinity. Here, $v^*$ indicates the baroclinic velocity and is defined as $v^* = v - \hat{v}$,
where $v$ is the meridional velocity and $\hat{v}$ the barotropic meridional velocity (i.e., the section spatially-averaged meridional velocity).
The quantity $\langle S \rangle$ indicates the zonally-averaged salinity and primed quantities ($v'$ and $S'$) are deviations from 
their respective zonal averages  \cite{Juling2021}. We will use  $F_{\mathrm{ovS}} = 
F_{\mathrm{ov}}(y = 34^\circ$S) and $F_{\mathrm{ovN}} = F_{\mathrm{ov}}(y = 60^\circ$N, 
with similar expressions for  $F_{\mathrm{azS}}$ and $F_{\mathrm{azN}})$, respectively.

\section{Results}

\subsection{AMOC Hysteresis}

\begin{figure}[htbp]

\center
\includegraphics[width=1\columnwidth]{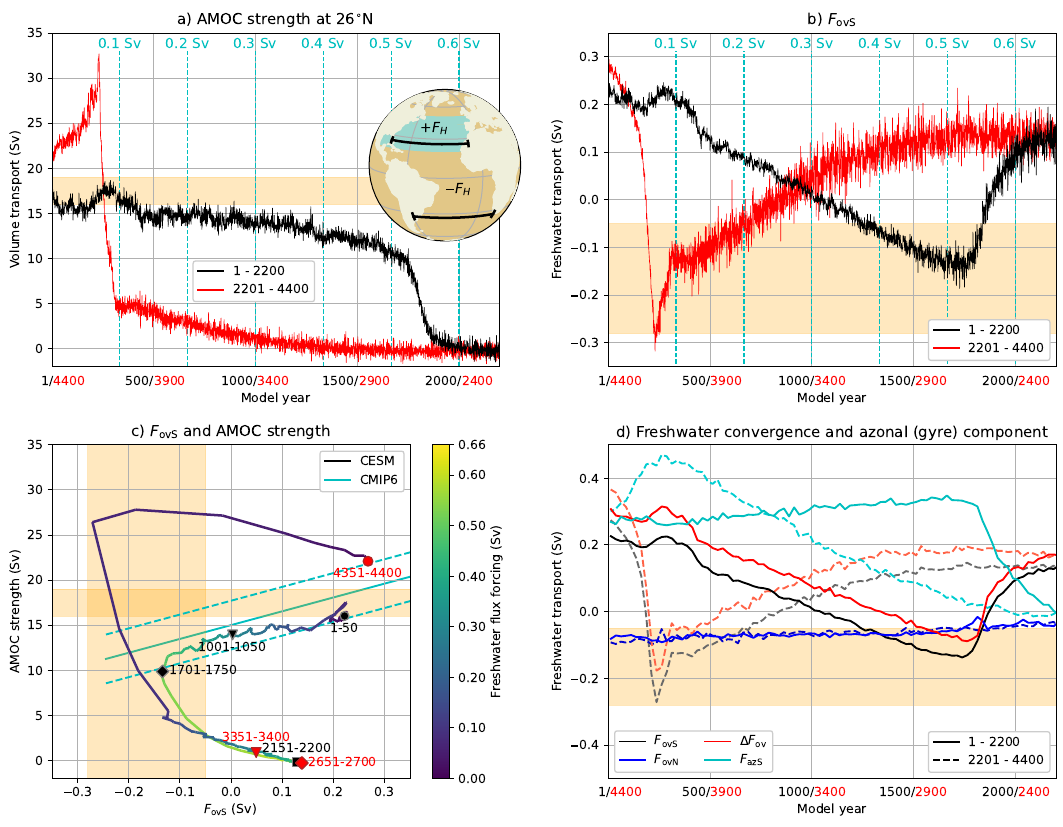}
\caption{(a): The AMOC strength at 1,000~m and 26$^{\circ}$N.
The cyan-coloured lines in panels~a and b indicate the magnitude of the freshwater forcing 
$F_H$. Inset: Fresh water is added to the ocean surface between 20$^{\circ}$N -- 50$^{\circ}$N 
in the Atlantic Ocean ($+F_H$)  and is compensated over the remaining ocean surface 
($-F_H$). The black sections indicate the 26$^{\circ}$N and 34$^{\circ}$S latitudes over 
which the AMOC strength and $F_{\mathrm{ovS}}$ are determined, respectively.
(b): The freshwater transport by the AMOC at 34$^{\circ}$S, $F_{\mathrm{ovS}}$.
(c): The AMOC strength versus  $F_{\mathrm{ovS}}$ with time parameterised; 
the time series are displayed as 25-year averages (to reduce their variability). 
The markers indicate the 50-year average over a particular period,
where the same marker shape indicates the same freshwater forcing ($F_H$) between two periods.
The cyan-coloured curve indicates the present-day (1994 -- 2020) CMIP6 regression and 1 standard deviation \cite{vanWesten2023b}.
(d): The freshwater transport at 34$^{\circ}$S ($F_{\mathrm{ovS}}$, black curve), 
60$^{\circ}$N ($F_{\mathrm{ovN}}$, blue curve), the freshwater convergence ($\Delta F_{\mathrm{ov}} = F_{\mathrm{ovS}} - F_{\mathrm{ovN}}$, red curve) 
and the azonal (gyre) component at 34$^{\circ}$S ($F_{\mathrm{azS}}$, cyan curve).
The time series are displayed as 25-year averages (to reduce the variability of the time series).
The yellow shading in all panels indicates observed ranges \cite{Garzoli2013, Mecking2017, Smeed2018, Worthington2021} for $F_{\mathrm{ovS}}$ and AMOC strength.
}
\label{fig:Figure_1}
\end{figure}

\begin{figure}[h!]

\includegraphics[width=1\columnwidth]{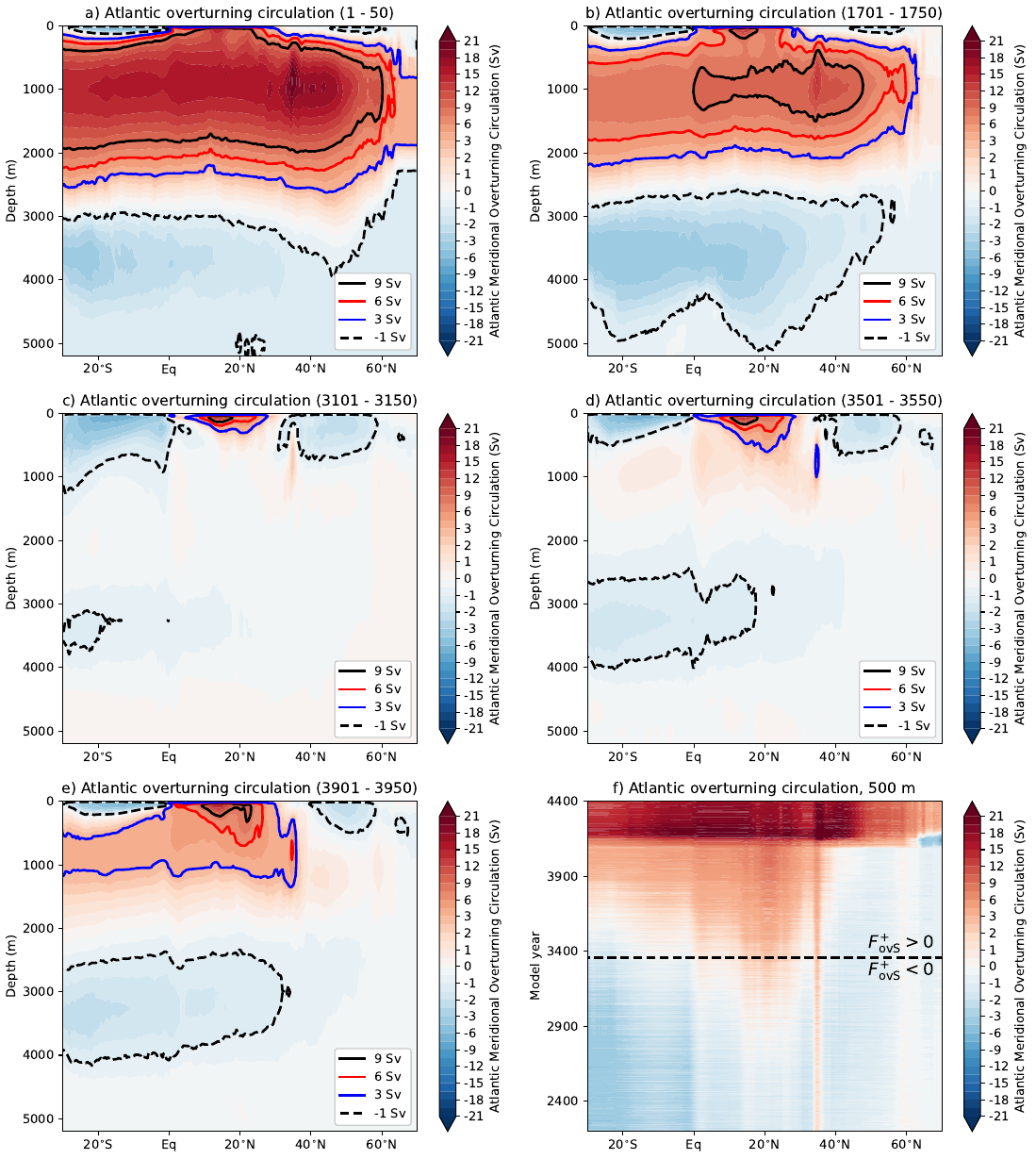}

\caption{(a -- e): The AMOC streamfunction ($\Psi$) for various 50-year periods.
The contours indicate the isolines  for different AMOC values as indicated in the legend. 
(f): Hovm\"oller diagram of the AMOC at 500~m depth for the reverse  simulation where 
the dashed line indicates where $F_{\mathrm{ovS}}$ switches sign in the forward 
simulation.}
\label{fig:Figure_2}
\end{figure}

The hysteresis behaviour for the AMOC strength is shown in Figure~\ref{fig:Figure_1}a,
where the black curve is the forward simulation (increasing freshwater flux forcing) and the 
red curve is the reversed simulation.  We display time on the horizontal axis as we follow a 
quasi-equilibrium approach and  the magnitude of the freshwater flux over time is indicated 
by the cyan-coloured dashed lines. The initial AMOC strength is about 17 Sv and the AMOC  pattern 
is shown in Figure~\ref{fig:Figure_2}a. In the forward simulation, the AMOC collapses  near model 
year~1,750 from about 10~Sv (Figure~\ref{fig:Figure_2}b) to about 2~Sv  (model year~1850) and 
it is near zero Sv  at model year~2,200. The freshwater transport carried by AMOC at 34$^{\circ}$S, 
$F_{\mathrm{ovS}}$, decreases in the forward experiment, goes through zero at model year~1,000 
and after the AMOC collapse it becomes positive (Figure~\ref{fig:Figure_1}b). The 
details of the AMOC collapse, its climate impact and its early warning signals were presented
elsewhere \cite{vanWesten2023c}.  

In the reversed simulation, years 2,200 to 4,400,  the AMOC stays near zero Sv to about model year~3,100 (Figure~\ref{fig:Figure_2}c).
A weak and shallow ($< 1000$~m) northward overturning cell develops 400~years later 
(Figure~\ref{fig:Figure_2}d) and the AMOC strength (at 1,000~m and 26$^{\circ}$N) is about 1.5~Sv.
This overturning cell strengthens over time (Figure~\ref{fig:Figure_2}e) and there is a clear northward 
overturning cell present prior to (full) AMOC recovery,  the AMOC strength (at 1,000~m and 26$^{\circ}$N) is about 4~Sv. 
This gradual AMOC increase of about 5~Sv over a 1,000-year period is 
followed by AMOC recovery between model years~4,090 and 4,170 (Figure~\ref{fig:Figure_1}a) 
where  its strength increases from about 5~Sv to a maximum of 33~Sv and thereafter slowly decreases again. 
Hence, there is a broad interval in  the freshwater 
flux forcing  ($F_H$ varies from about 0.1 to 0.5~Sv) which determines the hysteresis width (about 
0.4~Sv). In the reversed simulation,   $F_{\mathrm{ovS}}$ gradually decreases with decreasing  
freshwater flux   forcing   from a positive value at model year~2,200  to about  -0.12~Sv 
prior to AMOC recovery.  Then $F_{\mathrm{ovS}}$ steeply decreases to minimum of -0.32~Sv 
in model year~4,170 and afterwards strongly increases. 

It is interesting to compare the responses in the different freshwater transport components under the varying 
freshwater flux forcing   (Figure~\ref{fig:Figure_1}d) with those expected from  the bifurcation analyses from idealised 
ocean-climate model studies \cite{Dijkstra2007, Huisman2010}.   Based on these idealised model results, 
one expects that the AMOC has one steady state for $F_{\mathrm{ovS}}^+ > 0$, the + superscript indicating 
its value on  the forward simulation.  The multiple equilibrium regime would be entered by model year~1,000 
(where $F_{\mathrm{ovS}}$ changes sign) and hence AMOC recovery  would be expected near  
$F_H \sim  0.3$~Sv, which is around model year~3,400 in the reversed simulation.   Note that some delay in recovery can 
be expected due to  the quasi-equilibrium approach \cite{Hawkins2011}, but  the AMOC recovery 
in the CESM, is 700~years later than expected.  Hence, at first sight, this is substantially different 
from what  idealised model studies  \cite{Dijkstra2007, Huisman2010} would 
suggest.  

However, the Hovm\"oller diagram for AMOC at 500~m depth (Figure~\ref{fig:Figure_2}f) shows the development of 
the northward overturning cell (from 30$^{\circ}$S to 40$^{\circ}$N) from model year~3,500 and onwards; 
the Hovm\"oller diagrams between 500~m and 1,000~m depths show similar results (not shown).
The timing for the development of the northward overturning cell aligns well with the sign change in 
$F_{\mathrm{ovS}}^+$. Prior to the full AMOC recovery, the AMOC strength remains fairly low 
($< 5$~Sv) and the overturning cell has a maximum meridional 
extent to 40$^{\circ}$N (Figure~\ref{fig:Figure_2}f).  When the AMOC fully 
recovers (from model year 4,090), the northward overturning cell extends to higher latitudes 
(60$^{\circ}$N). This suggest that between the model years $3,400$ and $4,090$, there are 
processes  in the CESM that prevent a larger meridional extent of the overturning cell 
and delay a full AMOC recovery; details are discussed in the next subsection. 

The $F_{\mathrm{ovS}}$ minimum at model year~4,170 can be understood from the different adjustment time scales for 
the salinity and velocity fields at 34$^{\circ}$S (Figure~S1). The velocity fields adjust faster to the changing forcing 
than the salinity fields and velocity changes  then control the magnitude of $F_{\mathrm{ovS}}$. Prior to AMOC recovery, 
$F_{\mathrm{ovS}}$ is negative and hence the AMOC transports salt into the Atlantic basin. Surface velocities 
increase rapidly, as the AMOC increases during its recovery, and hence more salt is transported into the Atlantic 
basin, reducing the value of  $F_{\mathrm{ovS}}$.  Eventually,  the salinity fields adjust to the recovered 
ocean circulation state and $F_{\mathrm{ovS}}$ becomes positive again. The $F_{\mathrm{ovS}}$ value at 
the end of the simulation is slightly higher than its initial value,  because  the 34$^{\circ}$S 
salinity profiles are still different than the initial profiles. It is expected that AMOC and $F_{\mathrm{ovS}}$ 
eventually recover to their initial state (Figure~\ref{fig:Figure_1}c)  when extending the simulation 
beyond model year~4,400 under zero anomalous  freshwater flux forcing ($F_H = 0$~Sv), but we 
did not perform a simulation to check this. 

\begin{figure}[htbp]

\includegraphics[width=1\columnwidth]{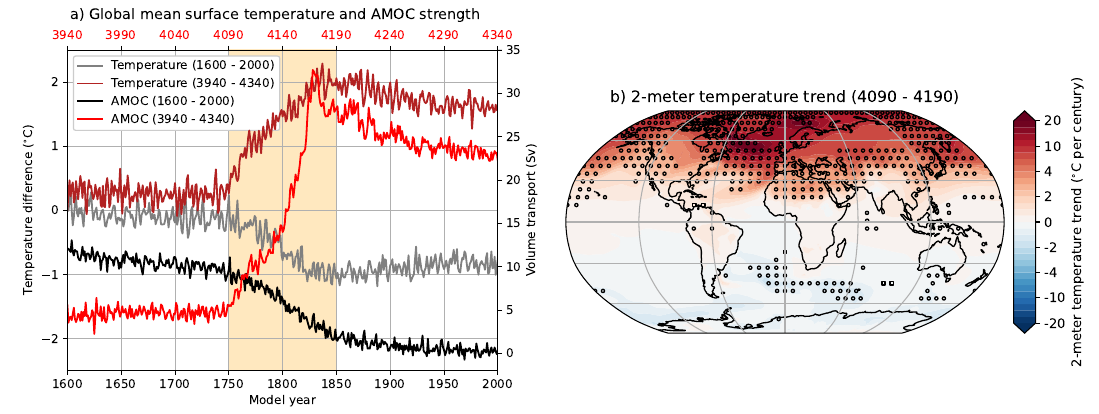}

\caption{(a): The globally-averaged 2-meter surface temperature and AMOC strength (at 1,000~m and 26$^{\circ}$N, similar as Figure~\ref{fig:Figure_1}a)
for the AMOC collapse (lower x-axis) and AMOC recovery (upper x-axis).
The temperature differences are w.r.t. model year~1,600 (grey curve) and model year~3,940 (brown curve).
The yellow shading indicates the 100-year period centred around the AMOC transient responses
and is used for determining the 2-meter temperature trends in panel~b.
(b): The 2-meter surface temperature trend (model years~4,090 -- 4,190) during AMOC recovery.
The circled (squared) markers indicate where the absolute magnitude of the trend are more than 50\% larger (smaller) during AMOC recovery 
than during AMOC collapse (model years~1,750 -- 1,850), and are only displayed when the local trend over both periods are significant ($p < 0.05$, two-sided t-test \cite{Santer2000}).}

\label{fig:Figure_3}
\end{figure}

During its recovery, the  AMOC strongly overshoots and its maximum (32.7~Sv in model year~4,168)  is about 
twice as strong compared to its initial value. This  overshoot is expected because meridional temperature and salinity 
gradients are (strongly) different when comparing those to the initial state (Figure~S2) and favour a (temporarily) 
stronger AMOC state. The transient response is also larger during AMOC recovery than during AMOC collapse 
(Figure~\ref{fig:Figure_3}a)  and when comparing two 100-year periods (model years~1,750 -- 1,850 versus 3,090 -- 3,190)
the (absolute) linear trends differ by a factor of 3.6 for AMOC strength. Over this 100-year period, however, the 
AMOC strength shows a non-linear change (in particular for the AMOC recovery) and  when determining the 
trends over a 20-year sliding window we find maximum (absolute) changes of 0.12~Sv~yr$^{-1}$ (AMOC 
collapse) and 0.76~Sv~yr$^{-1}$, which is about a factor 6~difference. 

The patterns in 2-meter atmospheric surface temperature trends are very similar between collapse and recovery 
(Figure~\ref{fig:Figure_3}b) but of course  differ by the sign of the trend. The magnitudes of the trends are, however,
larger during AMOC recovery (in particular for the Northern Hemisphere) and the globally-averaged trend is a factor~1.5 larger during AMOC recovery than during collapse. Note that,  during AMOC collapse   and recovery, 
the changes in freshwater  forcing $F_H$ are very small ($< 0.03$~Sv), meaning that the transient 
responses are mainly driven by internal feedbacks.

\subsection{AMOC Recovery: Role of sea ice}

\begin{figure}[h]

\includegraphics[width=1\columnwidth]{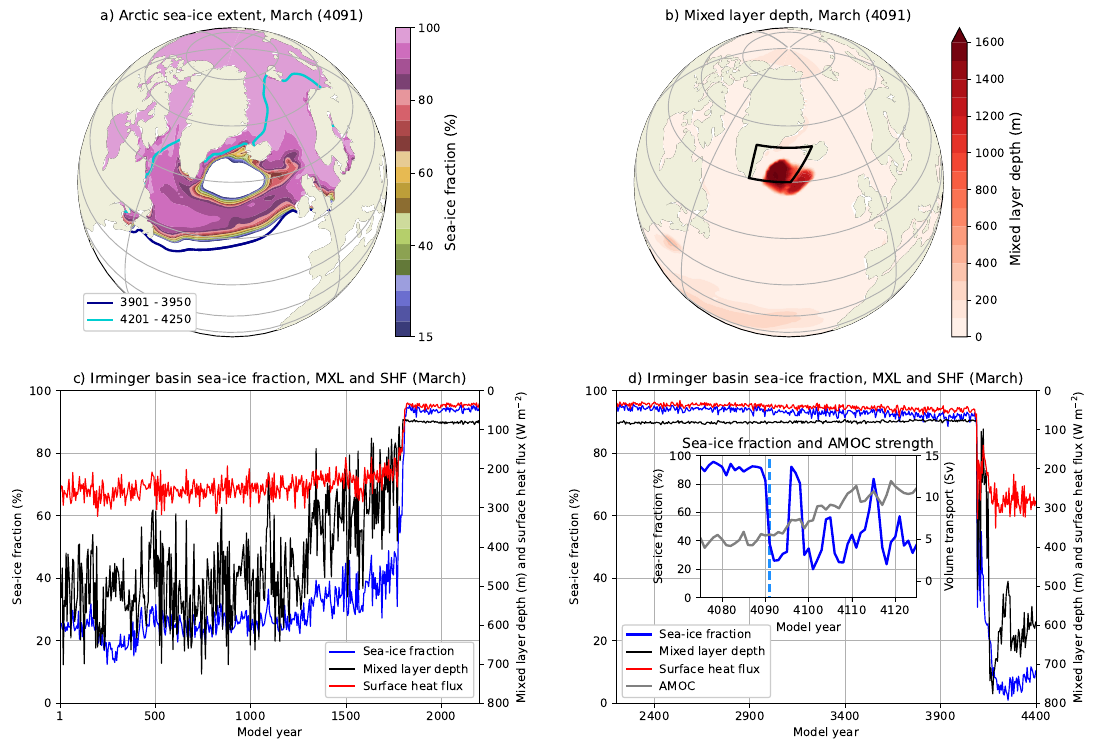}

\caption{(a): The sea-ice fraction for March model year~4,091, the curves display the sea-ice edge (i.e., the 15\% sea-ice fraction isoline) for model years~3,901 -- 3950 (dark blue curve) and 4,201 -- 4250 (cyan curve).
(b): The mixed layer depth for March model year~4,091. The black outlined region is the Irminger basin, used in panels~c and d.
(c \& d): The spatially-averaged time series of sea-ice fraction, mixed layer depth (MXL) and surface heat flux (SHF) over the Irminger basin for March. 
All values of SHF are positive indicating heat loss from the ocean to the atmosphere (note the flipped vertical range on the right axis).
The time series are displayed as 5-year averages (to reduce the variability of the time series).
The inset in panel~d shows the March sea-ice fraction (over the Irminger basin) and the yearly-averaged AMOC strength (at 1,000~m and 26$^{\circ}$N).  The dashed line indicates model year~4,091 when the (Irminger basin) polynya forms.}

\label{fig:Figure_4}
\end{figure}

To explain the delay in AMOC recovery, we consider the sea-ice distribution in the North Atlantic, which 
has undergone a vast expansion after the  AMOC collapse \cite{vanWesten2023c}.  From a recent  CMIP6 
model study  it was shown that sea ice suppresses air-sea fluxes and a relatively large sea-ice cover results 
in a general weaker upper ocean mixing and AMOC strength \cite{Lin2023}. In the CESM simulation here, 
this effect of sea ice on air-sea fluxes and upper ocean mixing also occurs over the Irminger basin (region 
of deep convection) during and after the AMOC collapse (Figure~\ref{fig:Figure_4}c). The sea-ice cover in 
the CESM is actually quite large compared to many CMIP6 models \cite{Lin2023} and explains the weak 
AMOC state ($< 5$~Sv). The sea-ice edge of the collapsed state extends down to 45$^{\circ}$N in the 
Atlantic Ocean (blue curve in Figure~\ref{fig:Figure_4}a)  and this restricts the meridional extent of the 
northward overturning cell to 40$^{\circ}$N (Figure~\ref{fig:Figure_2}d).

During the gradual spin-up of the northward overturning cell from model year~3,500 and onwards, 
more heat is transported northward leading to less favourable sea-ice formation conditions over time.
During the winter of model years~4,090/4,091, a few years before the full AMOC recovery, the seasonally-varying Arctic sea ice advances southward (Figure~S3).
The enhanced meridional heat transport prevents the Irminger basin from (fully) freezing over and a small polynya, 
i.e., a sea-ice free region (with sea-ice fractions lower than 15\%) within the sea-ice pack, forms. 
The sea-ice free ocean surface is strongly cooled by the atmosphere resulting in buoyancy loss and deep convection. 
Deep convection mixes relatively warm water from greater depth to the surface and this prevents the polynya from freezing over.
The polynya starts to grow in the following months and in March~model year~4,091 the polynya has a surface area of $6.46 \times 10^{5}$~km$^{2}$ (Figure~\ref{fig:Figure_4}a)
and the mixed layer depth extends down to 1,600~m (Figure~\ref{fig:Figure_4}b).

The next winter (model years~4,091/4,092) the ocean surface near the Irminger basin remains relatively warm (from the deep convection of the polynya). This prevents again a complete sea-ice coverage over the Irminger basin (inset in Figure~\ref{fig:Figure_4}d) 
and the sea-ice free region induces again deep convection.
These deep convection events during the winter months start to spin-up the AMOC 
and,  from model year~4,094,  the AMOC strength strongly increases (inset in Figure~\ref{fig:Figure_4}d).
The stronger AMOC results in more meridional heat transport leading to less sea-ice cover and more deep convection (Figure~\ref{fig:Figure_4}d)  which further increases the AMOC strength. 
Eventually the sea-ice cover retreats further northward (cyan curve in Figure~\ref{fig:Figure_4}a) and hardly 
influences deep convection anymore over the Irminger basin (and Labrador basin).

In summary, 
the strong effect of North Atlantic  sea-ice on the AMOC is responsible for the delay in the AMOC recovery. 
The  AMOC restarts  around model year~3,400  when $F_{\mathrm{ovS}}^{+}$ switches sign 
(Figure~\ref{fig:Figure_2}f), but remains weak due 
to the extensive sea-ice cover. Only when passing a critical value of the AMOC strength, which depends 
on the sea-ice distribution, the recovery is completed.  The Irminger basin polynya during the winter of model 
years~4,090/4,091 marks the point in time when AMOC recovery is inevitable.  This explains why such 
sea-ice induced AMOC recovery delay is not found  in models where there is no sea-ice component  
\cite{Rahmstorf1996, Dijkstra2007, Huisman2010}. It also implies that in CMIP6 models, which 
have a strong sea-ice model  dependency \cite{Lin2023},  it is expected that the timing of AMOC 
recovery is quite  model dependent.

\section{Discussion}

The  Atlantic Meridional Overturning Circulation (AMOC) is considered to be a tipping element 
in the climate system \cite{Armstrong2022}. We have shown here a first full AMOC hysteresis 
simulation in the Community Earth System Model, a state-of-the-art Global Climate Model.  This 
result indicates that  AMOC tipping does not only occur in conceptual \cite{Stommel1961} and 
idealised  climate  models \cite{Rahmstorf2005}, but actually  in the full hierarchy of climate models. 
The shape of the hysteresis is also relatively smooth and without sharp strong local transitions as 
occur in ocean-only models \cite{Rahmstorf1996, Lohmann2023}. 

The  hysteresis width, measured as the difference in freshwater forcing between 
collapse and recovery,  found in the CESM (about 0.4~Sv) is quite large compared to that 
found in many  EMICs  \cite{Rahmstorf2005}, also much larger than that in CCSM 
\cite{Hu2012}, but similar to that in FAMOUS \cite{Hawkins2011}.   We have 
shown here that the Northern Hemisphere sea-ice distribution has a substantial effect on 
the recovery of the AMOC. The fact that sea ice inhibits air-sea fluxes, and hence convection,  
appears important  \cite{Lin2023} for the recovery of the AMOC in the CESM and hence 
for the hysteresis width. The sea ice
is  the main reason for the strong asymmetry in the hysteresis, where the recovery is 
a factor six faster than the collapse. 
This strong asymmetry is likely relevant for explaining proxy records  of the Dansgaard-Oeschger 
events \cite{Rahmstorf2002, Henry2016, Lynch-Stieglitz2017}. These events clearly show that warming is 
much faster than the cooling, consistent with that AMOC recovery occurs on (much) shorter time 
scales than its collapse. 

What is  intriguing  is that the AMOC collapse \cite{vanWesten2023c} and recovery can be 
connected  to theory about the role of  the AMOC induced freshwater transport at the southern boundary 
of the Atlantic, $F_{\mathrm{ovS}}$, on AMOC stability.   In  idealised  global ocean models, the 
recovery tipping point  is connected to a sign change in  $F_{\mathrm{ovS}}^+$ on the northward
overturning  AMOC  state. At this point,  a change from the AMOC exporting salt to 
exporting freshwater occurs \cite{Dijkstra2007}. In the CESM simulation here, the 
AMOC indeed restarts when  there is this sign change in $F_{\mathrm{ovS}}$, but the AMOC strength 
remains low due to the effect of the sea ice. These results indicate  that the AMOC hysteresis can be 
captured in a low-dimensional dynamical context giving more  weight to the results of conceptual 
and idealized AMOC models.  

The CESM version used here has  substantial  biases in freshwater forcing in its  pre-industrial
reference  state \cite{vanWesten2023c}, which is the reason that the initial
value of $F_{\mathrm{ovS}}$ is positive.  It has been shown using  the global 
ocean model as used in \citeA{Dijkstra2007} that this shifts the AMOC tipping points 
associated with collapse and recovery to higher values of the freshwater forcing 
\cite{Dijkstra2023}.   Assuming that the CESM will show a similar qualitative behaviour, this 
implies that  collapse and recovery, when biases are corrected, will 
likely occur for much smaller values of the freshwater forcing than used in the simulation 
here. 
Combined with the change from pre-industrial to present-day 
forcing, and including  climate change, it remains to be investigated
whether the AMOC can tip before the year~2100  \cite{Ditlevsen2023} and assess the  
probability of such a transition. The results  here have demonstrated that this is 
an  important  line of research as such tipping behaviour occurs in models which are 
considered to adequately represent the development of Earth's large-scale climate. 

\section*{Open Research Section}
All model output and code to generate the results are available 
\cite[]{vanWesten2023d} from  \sloppy{https://doi.org/10.5281/zenodo.8262424}. 

\acknowledgments
We thank Michael Kliphuis (IMAU, UU) for  performing  the CESM simulation.
The model simulation and the analysis of all the model output was conducted on 
the Dutch National  Supercomputer (Snellius) within NWO-SURF project~17239. 
R.M.v.W. and H.A.D. are funded by the European Research Council through the 
ERC-AdG project TAOC (project~101055096).  

%
%


\bibliography{References,erc_2021_total}

%
%
%
%
%

\renewcommand{\thefigure}{S\arabic{figure}}
\setcounter{figure}{0}   

\begin{figure}[h]

\hspace{-3.5cm}
\includegraphics[width=1.5\columnwidth]{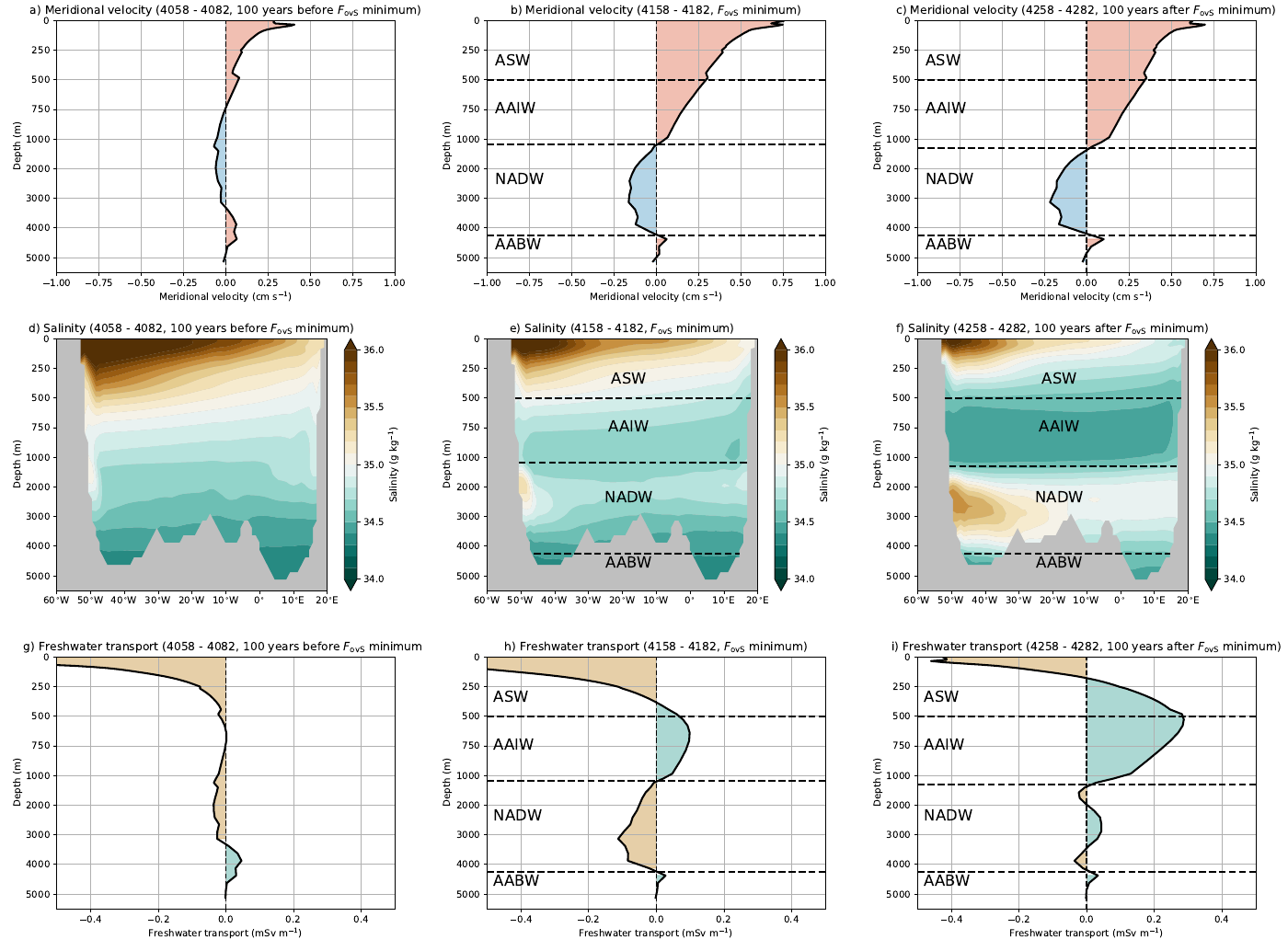}

\caption{(Upper row): The zonally-averaged meridional velocity at 34$^{\circ}$S for three 25-year periods:
100~years before $F_{\mathrm{ovS}}$ minimum, at the $F_{\mathrm{ovS}}$ minimum and 100~years after the $F_{\mathrm{ovS}}$.
(Middle row): The salinity along 34$^{\circ}$S for the three periods.
(Lower row): The freshwater transport with depth at 34$^{\circ}$S for the three periods.
The different water masses are derived from the velocity profile \cite{vanWesten2023b} and is only applicable for the (strong) northward overturning circulation (middle and right column) and
the names are: Atlantic Surface Water (ASW), Antarctic Intermediate Water (AAIW), North Atlantic Deep Water (NADW) and Antarctic Bottom Water (AABW).}

\label{fig:Figure_S1}
\end{figure}


\begin{figure}[h]

\hspace{-3.5cm}
\includegraphics[width=1.5\columnwidth]{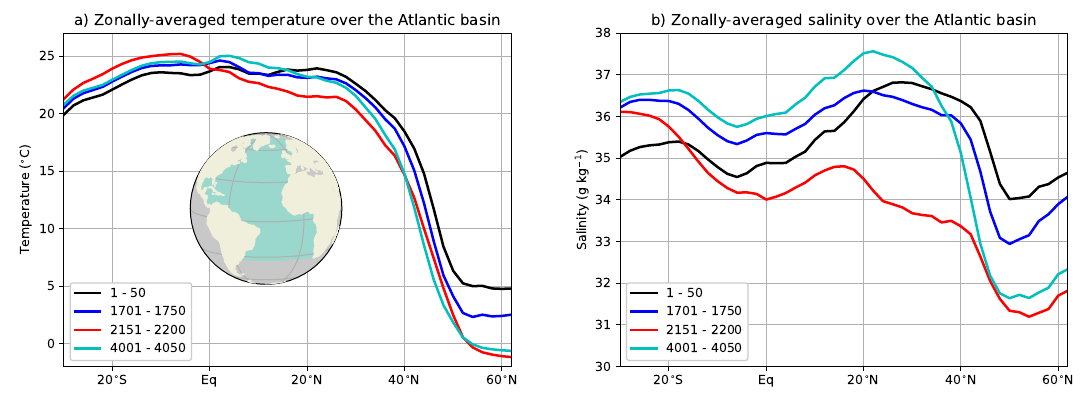}
\caption{The volume-averaged (upper 100~m) and zonally-averaged (a): temperature and (b): salinity over the Atlantic basin (blue region in inset in panel~a) for
4 different 50-year periods.}
\label{fig:Figure_S2}
\end{figure}


\begin{figure}

\hspace{-3.5cm}
\includegraphics[width=1.5\columnwidth]{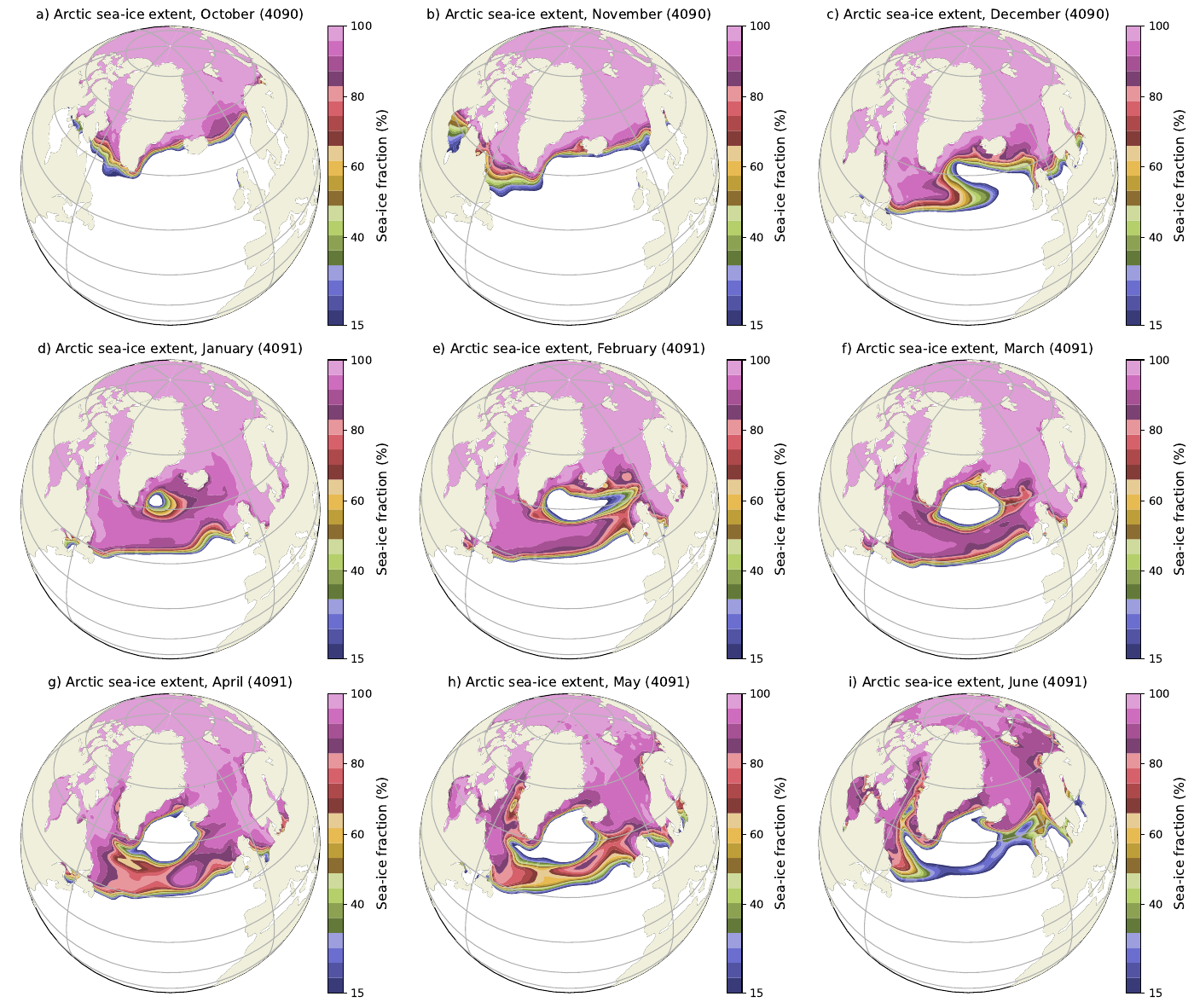}
\caption{Sea-ice fractions between October model year~4,090 and June model year~4,091.}
\label{fig:Figure_S3}
\end{figure}

\end{document}